\documentclass[aps,showpacs,amsmath,amssymb,twocolumn,nofootinbib]{revtex4-1}

\usepackage{cancel}
\usepackage{cases}
\usepackage{subfigure}
\usepackage{graphicx}% Include figure files
\usepackage{dcolumn}% Align table columns on decimal point
\usepackage{bm}% bold math
\usepackage{color}
\usepackage[colorlinks=true,pdfstartview=FitV,linkcolor=blue,citecolor=blue,urlcolor=blue]{hyperref}
\usepackage[mathlines]{lineno}% Enable numbering of text and display math
\usepackage[dvipsnames]{xcolor}
\usepackage{amsmath}
\usepackage{ytableau}

\everymath{\displaystyle}
\begin{document}

\newcommand{\beq}{\begin{equation}}
\newcommand{\eeq}{\end{equation}}
\newcommand{\beqs}{\begin{eqnarray}}
\newcommand{\eeqs}{\end{eqnarray}}
\newcommand{\rbf}{\color{red}}

\title{Charms of Strongly Interacting Conformal Gauge Mediation}

\author{Gongjun Choi,$^{1}$}
\thanks{{\color{blue}gongjun.choi@gmail.com}}

\author{Tsutomu T. Yanagida,$^{1,2}$}
\thanks{{\color{blue}tsutomu.tyanagida@sjtu.edu}}

\author{Norimi Yokozaki,$^{3}$}
\thanks{{\color{blue}n.yokozaki@gmail.com}}

\affiliation{$^{1}$ Tsung-Dao Lee 
Institute, Shanghai Jiao Tong University, Shanghai 200240, China}

\affiliation{$^{2}$ Kavli IPMU (WPI), UTIAS, The University of Tokyo,
5-1-5 Kashiwanoha, Kashiwa, Chiba 277-8583, Japan}

\affiliation{$^{3}$ Zhejiang Institute of Modern Physics and Department of Physics, Zhejiang University, Hangzhou, Zhejiang 310027, China}
\date{\today}

\begin{abstract}
By extending a previously proposed conformal gauge mediation model, we construct a gauge-mediated SUSY breaking (GMSB) model where a SUSY-breaking scale, a messenger mass, the $\mu$-parameter and the gravitino mass in a minimal supersymmetric (SUSY) Standard Model (MSSM) are all explained by a single mass scale, a R-symmetry breaking scale. We focus on a low scale SUSY-breaking scenario with the gravitino mass $m_{3/2}=\mathcal{O}(1){\rm eV}$, which is free from the cosmological gravitino problem and relaxes the fine-tuning of the cosmological constant. Both the messenger and SUSY-breaking sectors are subject to a hidden strong dynamics with the conformality above the messenger mass threshold (and hence the name of the model “strongly interacting conformal gauge mediation”). In our model, the Higgs B-term is suppressed and a large $\tan\beta$ is predicted, resulting in the relatively light second CP-even Higgs and the CP-odd Higgs with a sizable production cross section. These Higgs bosons can be tested at future LHC experiments. 
\end{abstract}

\maketitle
\section{Introduction}  
Field theories containing a scalar field with an intermediate mass scale have a common pathology even if they are renormalizable; for a quantum correction to the scalar mass squared ($m_{\rm scalar}^{2}$), there appears the quadratic divergence proportional to $\Lambda_{\rm UV}^{2}$ where $\Lambda_{\rm UV}$ is a UV-cutoff. Notoriously, the Standard Model (SM) suffers from such a pathology due to the presence of the Higgs boson. However, imposing supersymmetry (SUSY) can cure this pathology in an elegant manner by introducing balance between bosonic and fermionic degrees of freedom in their contributions to $m_{\rm scalar}^{2}$. In addition, concerning the renormalization procedure of a theory, one needs only wave function renormalization for individual fields in SUSY theories. Thus, supersymmetrized theories are theoretically advantageous as compared to theories without SUSY as far as the issue of softening or removing divergences is concerned.

In spite of these merits that one can enjoy in a SUSY theory, there remain phenomenological problems in the minimal supersymmetric standard model (MSSM) such as the flavor changing neutral current (FCNC) and the CP problem~\cite{Misiak:1997ei,Dine:1996ui,Gabbiani:1996hi}. In addition, the observed vanishingly small cosmological constant demands severe fine-tuned cancellation between a F-term SUSY-breaking contribution and a R-symmetry breaking contribution to a scalar potential. As for these difficulties, the low scale gauge mediated SUSY-breaking (GMSB) scenario seems to be almost a unique solution~\cite{Dine:1993yw,Dine:1994vc,Dine:1995ag}; a gauge mediation model can escape from the former two problems. And the lower a SUSY-breaking scale becomes, the milder fine-tuning for the observed vanishingly small cosmological constant is required.

Bearing these merits of GMSB models, in this paper, we give our special attention to a class of GMSB models dubbed “conformal gauge mediation (CGM)”~\cite{Ibe:2007wp,Ibe:2008si} (see also the follow-up works in~\cite{Yanagida:2010wf,Yanagida:2010zz}). Differing from many of GMSB models, this type of GMSB model has the interesting property of having less number of free parameters. To be more specific, the model has a messenger mass as a single fundamental parameter which determines a SUSY-breaking scale and SUSY particle mass spectrum. Above the messenger mass scale, gauge coupling and Yukawa coupling constants are assumed to be near infrared fixed points (IRFP) and hence the name “conformal gauge mediation”.

In this work, we construct and study an extended strongly interacting conformal gauge mediation model by introducing a new $R$-symmetry breaking chiral superfield $\Phi$. The new ingredient of our model as compared to the previous CGM is that an energy scale of $R$-symmetry breaking serves as the fundamental single parameter of the model. Thereby, we attribute not only a SUSY-breaking and messenger mass scale but also R-symmetry breaking scale to a common origin. For the purpose of softening the degree of fine-tuning in producing the observed tiny cosmological constant, we consider the scenario with the gravitino mass around $m_{3/2}=\mathcal{O}(1){\rm eV}$. The gravitino in this mass range is also favored from the view point of cosmology: we do not have an upper bound on the reheating temperature~\cite{Moroi:1993mb}, and hence, the thermal leptogenesis~\cite{Fukugita:1986hr} can explain the observed baryon asymmetry in a naive way. We shall see that the model is featured by the R-symmetry breaking scale near $10^{11}{\rm GeV}$, which in turn provides us with the messenger mass and the SUSY-breaking scale amounting to $\mathcal{O}(100){\rm TeV}$ and $\mathcal{O}(1){\rm TeV}$ Higgsino mass .

In Sec.~\ref{sec:model}, we first make a brief review for the existing primordial form of the CGM model. After that, we discuss the key essence of our extended model which is the introduction of a chiral superfield $\Phi$ breaking R-symmetry in the model in the spontaneous way and how we infer the R-symmetry breaking scale. In Sec.~\ref{sec:result}, we discuss several interesting consequences and features of the model including how the model can give us a natural explanation for $\mathcal{O}(100){\rm TeV}$ messenger mass, how the model generates $\mu$ and $B$-terms, how the model helps us understand the electroweak scale, and why the introduction of $\Phi$ is harmless in cosmology and the dark matter candidate of the model. Finally, in Sec.~\ref{sec:conclusion}, we conclude this paper by summarizing the structure and resultant aspects of the model.

\section{Strongly interacting conformal gauge mediation model}\label{sec:model}
\subsection{Hidden strong dynamics}\label{sec:hiddenstrong}
In a supergravity model, the scalar potential at the leading order is determined by a difference between $|F|^{2}$ and $3W_{0}^{2}/M_{P}^{2}$ where $F$ is the order parameter for the SUSY-breaking, $W_{0}$ is R symmetry-breaking constant term in a superpotential and $M_{P}\simeq2.4\times10^{18}{\rm GeV}$ is the reduced Planck mass. Now that the observed energy density of cosmological constant turns out to be negligibly tiny, we expect $|F|\simeq\sqrt{3}W_{0}/M_{P}=\sqrt{3}m_{3/2}M_{P}$. Given this, it is realized that the smaller gravitino mass makes the degree of fine-tuned cancellation between two contributions to the scalar potential less severe. The desire to avoid unnatural miraculous fine-tuning in parameters in the model prefers lower values of $|F|$ and $m_{3/2}$, posing a question: how low $m_{3/2}$ could be?

In \cite{Osato:2016ixc}, the stringent upper bound of the gravitino mass $m_{3/2}<4.7{\rm eV}$ ($95\%$C.L.) was reported based on the use of the data of CMB lensing and the cosmic shear observation.\footnote{See also Ref.~\cite{Viel:2005qj} where the matter power spectrum resulting from the Lyman-$\alpha$ forest data was used to obtain the weaker constraint $m_{3/2}<16{\rm eV}$ ($95\%$C.L.) with the energy density fraction $\omega_{3/2}/\omega_{\rm CDM}\lesssim0.12$.} On the other hand, $|F|$ being directly related to $m_{3/2}$, we expect that $m_{3/2}$ cannot be arbitrarily small to be consistent with the null observation of any SUSY particles in the LHC to date.\footnote{For instance, one may refer to Ref.~\cite{Aad:2014mra} where an $\mathcal{O}(1){\rm eV}$ lower bound on $m_{3/2}$ is found in the direct SUSY searches in the LHC.} Along this line of reasoning, we consider a scenario with $m_{3/2}=\mathcal{O}(1){\rm eV}$ corresponding to a SUSY-breaking scale $\sim\mathcal{O}(100){\rm TeV}$. 

Such a low SUSY-breaking scale might imply a GMSB scenario. Yet, we need $m_{3/2}\gtrsim\mathcal{O}(100){\rm eV}$ to explain the observed Higgs boson mass $125{\rm GeV}$ in a perturbative gauge mediation model~\cite{Ajaib:2012vc,Yanagida:2012ef}.  Therefore, it becomes necessary to consider the strongly coupled gauge mediation scenario~\cite{Yanagida:2012ef}. As an exemplary model satisfying this feature, we give our special attention to the strongly interacting conformal gauge mediation model~\cite{Ibe:2007wp,Ibe:2008si,Yanagida:2010wf,Yanagida:2010zz}. Therein SUSY-breaking is induced by the presence of a conformal phase of a hidden non-Abelian gauge theory of which the strong dynamics accounts for the interactions in the messenger sector and SUSY-breaking sector. Below, we first go through a review of the existing strongly interacting conformal gauge mediation model~\cite{Ibe:2007wp,Ibe:2008si,Yanagida:2010wf,Yanagida:2010zz} and then in the next subsection we extend it by introducing a new field content. 

For the promised strong interactions in the SUSY-breaking and messenger sectors, we consider $SU(N)$ hidden gauge group with $N_{Q}$ pairs ($i,\tilde{i}\!=\!1\!-\!N_{Q}$) of $(Q_{i},\tilde{Q}_{\tilde{i}})$ and $N_{P}$ pairs of ($a,\tilde{a}\!=\!1\!-\!N_{P}$) of $(P_{a},\tilde{P}_{a})$~\cite{Yanagida:2010wf,Yanagida:2010zz}. Here $(Q_{i},P_{a})$ and $(\tilde{Q}_{\tilde{i}},\tilde{P}_{a})$ are assumed to transform as the fundamental and anti-fundamental representations under $SU(N)$ respectively. Aside from these matter fields, we introduce $N_{Q}^{2}$ singlet fields $S_{i\tilde{i}}$ which couple to $(Q_{i},\tilde{Q}_{\tilde{i}})$. When a choice of $(N,N_{Q},N_{P})$ satisfies the condition $(3N)/2<N_{Q}+N_{P}<3N$, there exists an IRFP~\cite{Seiberg:1994pq}. For our purpose, we focus on the case with $(N,N_{Q},N_{P})=(4,3,5)$ by which the theory lies at the conformal window.

With these field contents, now we consider the following superpotential
\beq
W=\lambda S_{i\tilde{i}}Q_{i}\tilde{Q}_{\tilde{i}}+m_{P}P_{a}\tilde{P}_{a}\,,
\label{eq:superpotential1}
\eeq
where $m_{P}$ is a mass of $P_{a}$ and $\tilde{P}_{a}$, and $\lambda$ is a dimensionless coupling. $Q_{i}$ and $\tilde{Q}_{\tilde{i}}$ shall become massive once $S_{i\tilde{i}}$ acquires a vacuum expectation value (VEV). The massive matter fields $P_{a}$ and $\tilde{P}_{a}$ are taken to serve as messengers in the model. $P_{a}$ and $\tilde{P}_{a}$ respect the gauged flavor $SU(5)_{F}$ symmetry which we identify with $SU(5)_{\rm GUT}$ in the SM. 

We assume that the theory is in the close proximity of an IRFP of $SU(4)$ at the energy scale above $m_{P}$ and the scalar component of $S_{i\tilde{i}}$ acquires a non-zero VEV. For the simplicity of discussion hereafter, we take $S_{i\tilde{i}}$ to be diagonal in the flavor space, i.e. $S_{i\tilde{i}}=S\delta_{i\tilde{i}}$. Once the massive $(Q_{i},\tilde{Q}_{\tilde{i}})$ and $(P_{a},\tilde{P}_{a})$ are integrated-out, $S_{i\tilde{i}}$-dependence of the effective superpotential induced by the gaugino condensation reads~\cite{Yanagida:2010wf,Yanagida:2010zz}
\beq
W_{\rm eff}\sim{\rm det}[S_{i\tilde{i}}]^{\frac{1}{N}}=S^{\frac{N_{Q}}{N}}\,.
\label{eq:Weff}
\eeq

On the other hand, since the theory has an approximate superconformal symmetry at an energy scale higher than $m_{P}$, a low energy Kahler potential of the model is expected to have the term $\sim(S^{\dagger}S)^{1/\Delta_{S}}$ together with other $m_{P}$-dependent terms where $\Delta_{S}$ is a scaling dimension of $S$. Along with the superpotential in Eq.~(\ref{eq:Weff}), this term in a Kahler potential yields the scalar potential of the following form
\beq
V_{\rm scalar}\sim(S^{\dagger}S)^{\frac{\Delta_{S}-1}{\Delta_{S}}-1+\frac{N_{Q}}{N}}\,,
\label{eq:Vscalar}
\eeq
where we used the notation $S$ here to denote the scalar component of the chiral superfield $S$.
We can infer from Eq.~(\ref{eq:Vscalar}) that $V_{\rm scalar}$ can be stabilized provided $\Delta_{S}>N/N_{Q}$ holds. For $S$-field, the scaling dimension and the anomalous dimension at the IRFP ($\gamma_{S*}$) are related via $\Delta_{S}=1+\gamma_{S*}$ and $\gamma_{S*}$ is found to be $\sim0.5$ for $(N,N_{Q},N_{P})=(4,3,5)$~\cite{Izawa:2009nz,Yanagida:2010wf}. Thus we see that $\Delta_{S}>N/N_{Q}$ is indeed satisfied in the model, guaranteeing the spontaneous SUSY-breaking. Notice that the model could have IRFP thanks to the presence of $(P_{a},\tilde{P}_{a})$ fields, which is the essential point allowing for the SUSY-breaking.

By solving the coupled equations $\beta_{g_{4}}=\beta_{\lambda}=0$ with $\beta_{g_{4}}$ and $\beta_{\lambda}$ beta functions of $g_{4}$ and $\lambda$ respectively, IRFPs in the coupling space $(g_{4},\lambda)$ can be found and they read $(g_{4*},\lambda_{*})\simeq(3.6,3.2)$ at the one-loop level~\cite{Izawa:2009nz,Yanagida:2010wf}.\footnote{Here one loop level means that one-loop anomalous dimensions of fields appearing in Eq.~(\ref{eq:superpotential1}) and one-loop $\beta_{\lambda}$ were used together with NSVZ beta function for $g_{4}$.} Using this $g_{4*}$, we can obtain the relation between $m_{P}$ and an energy scale $\Lambda$ at which $g_{4}$ blows up via the following dimensional transmutation
\beq
\frac{8\pi^{2}}{g_{4*}^{2}}\simeq b_{1}\ln\left(\frac{m_{P}}{\Lambda}\right)\quad\rightarrow\quad m_{P}\sim\Lambda\,,
\label{eq:oneparameter}
\eeq
where $b_{1}=3N-N_{Q}=9$ is the one-loop coefficient of $\beta_{g_{4}}$ after $(P_{a},\tilde{P}_{a})$ are integrated-out. The empirical relation $\Lambda_{\rm QCD}\sim4\pi f_{\pi}$ between the dynamical (composite) scale $\Lambda_{\rm QCD}$ of QCD and the pion decay constant $f_{\pi}$ can provide us with the rough estimate of the dynamical scale $\Lambda_{4}$ of $SU(4)$ in our model in terms of $\Lambda$ in Eq.~(\ref{eq:oneparameter}), i.e. $\Lambda_{4}\sim4\pi\Lambda$. We notice, however, that there can be ambiguity in the factor $4\pi$ in the relation $\Lambda_{4}\sim4\pi\Lambda$ since our strong dynamics is different from QCD in the number of colors and matter fields, and exploring this factor rigorously is beyond the scope of the current paper. So whatever the factor is, the greater one among $m_{P}$ and $\Lambda_{4}$ is to be used for estimating the soft masses based on the naive dimensional analysis (NDA)~\cite{Luty:1997fk,Cohen:1997rt}. Under this circumstance, we assume $m_{P}\gtrsim\Lambda_{4}$ from here on. Now along this line of reasoning, Eq.~(\ref{eq:oneparameter}) informs us that the model has the non-trivial feature that a messenger mass equals the dynamical scale up to an $\mathcal{O}(1)$ factor and thereby it becomes the one parameter theory. With the SUSY-breaking induced dynamically, we infer the SUSY-breaking scale $F_{S}\sim \Lambda_{4}^{2}/(4\pi)$ where $F_{S}$ is the SUSY-breaking order parameter ($F$-term of the chiral superfield $S$)~\cite{Luty:1997fk,Cohen:1997rt}. 

Guided by an effective operator analysis~\cite{Ibe:2007wp,Ibe:2008si} and NDA, one can try to make an estimate for the sfermion and gaugino masses. Using $\lambda S\sim\Lambda_{4}+\Lambda_{4}^{2}\theta^{2}$ obtained from NDA and parametrizing the ambiguity arising from the strong dynamics in the hidden sector by $N_{\rm eff}$ and $N_{\rm eff}^{'}$ for each of sfermion and gaugino masses, the soft SUSY-breaking masses can be expressed as~\cite{Yanagida:2010wf,Yanagida:2010zz}
\beqs
&&m_{\rm sfermion}^{2} \sim 2N_{\rm eff}C_{2,{\rm SM}}({\ytableausetup{textmode, centertableaux, boxsize=0.6em}
\begin{ytableau}
 \\
\end{ytableau}} )\left(\frac{\alpha_{{\rm SM}}}{4\pi}\right)^{2}\frac{\Lambda_4^{4}}{m_{P}^{2}}\,,\cr\cr
&&m_{\rm gaugino}\sim N_{\rm eff}^{'}\left(\frac{\alpha_{{\rm SM}}}{4\pi}\right)\frac{\Lambda_4^{7}}{m_{P}^{6}}\,,
\eeqs
where $C_{2,{\rm SM}}({\ytableausetup{textmode, centertableaux, boxsize=0.6em}
\begin{ytableau}
 \\
\end{ytableau}})$ is a quadratic Casimir of a SM gauge group, $\alpha_{\rm SM}\equiv g_{\rm SM}^{2}/(4\pi)$ with $g_{\rm SM}$ a gauge coupling of a SM gauge group. Now that the relation between $\Lambda_{4}$ and $m_{P}$ is uncertain in our model, the ratio $\Lambda_{4}/m_{P}$ is treated as a free parameter. For the later discussion in Sec.~\ref{sec:muB}, we take a scale of the gaugino mass as a free parameter, imposing a GUT relation among the gaugino masses.

Recalling that SUSY-breaking scale around $\mathcal{O}(100){\rm TeV}$ is considered in the model, now we encounter a fundamental question for an origin of $m_{P}$. Namely, $m_{P}=\mathcal{O}(100){\rm TeV}$ needs an explanation for why it is much smaller than the Planck scale as a mass parameter of the vector-like fields $(P_{a},\tilde{P}_{a})$.

\subsection{R-symmetry breaking field $\Phi$}\label{sec:Phifield}
In this subsection, we show that we can answer the question concerning an origin of $m_{P}$ and its magnitude $\mathcal{O}(100){\rm TeV}$ by relying on a $R$-symmetry breaking field. What we discuss below is not the unique answer to the question. Nevertheless, it may lead on to non-trivial aspects of the model as explained in the next section, which motivates our exploration for the possibility presented below.

In any supergravity model where a SUSY-breaking scale deviates from the Planck scale, R-symmetry must be assumed in order to prevent $W_{0}$ from being $M_{P}^{3}$ and to reproduce the vanishingly small cosmological constant. With that being said, the gauged discrete $Z_{6R}$ is a particularly interesting R-symmetry because it satisfies the anomaly-free conditions for $Z_{NR}\otimes SU(2)_{L}^{2}$ and $Z_{NR}\otimes SU(3)_{c}^{2}$ in the MSSM with three families of quarks and leptons~\cite{Evans:2011mf}, and has $Z_{2R}$ as a subgroup to hinder the proton decay at the renormalizable level. Furthermore, once $Z_{6R}$ is assumed, the model can benefit from the automatic suppression of the dangerous dimension 5 proton decay operator {\bf 10}\,{\bf 10}\,{\bf 10}\,{\bf 5}$^{*}$ ~\cite{Sakai:1981pk,Weinberg:1981wj} and an intermediate scale Higgsino mass in a natural manner.\footnote{From the anomaly-free conditions for the mixed anomalies of $Z_{NR}\otimes SU(2)_{L}^{2}$ and $Z_{NR}\otimes SU(3)_{c}^{2}$ ($N\in\mathbb{N}$), and constraints from Yukawa interactions, it can be shown that R-charge of operators {\bf 10}\,{\bf 10}\,{\bf 10}\,{\bf 5}$^{*}$ and $H_{u}H_{d}$ are 0 modulo $N$ and 4 modulo $N$ respectively.} Motivated by this winning attributes, in our model, we impose the discrete gauged $Z_{6R}$ as a way to control the constant term in the superpotential. In Table.~\ref{table:qn}, we show R-charge assignment assumed in our model that makes the mixed anomalies of $Z_{6R}\otimes SU(2)_{L}^{2}$ and $Z_{6R}\otimes SU(3)_{c}^{2}$ free and Yukawa interactions in the MSSM respect $Z_{6R}$.

Above all, for $Z_{6R}$ to be anomaly-free with respect to $SU(4)$, R-charges of $(P_{a},\tilde{P}_{a})$ fields denoted by $R[P]$ and $R[\tilde{P}]$ should satisfy~\cite{Ibanez:1991hv,Ibanez:1991pr,Ibanez:1992ji}
\beq
4+\frac{5}{2}(R[P]+R[\tilde{P}]-2)=\frac{6}{2}k\qquad({\rm k\in\mathbb{Z}})\,.
\label{eq:Rcharge}
\eeq
Note that the mixed anomaly of $Z_{6R}\otimes SU(4)^{2}$ does not constrain R-charges of $(Q_{i},\tilde{Q}_{\tilde{i}})$ fields since $N_{Q}=3$. So we omitted the contribution of $(Q_{i},\tilde{Q}_{\tilde{i}})$ in Eq.~(\ref{eq:Rcharge}).\footnote{For $S_{i\tilde{i}}$ and $(Q_{i},\tilde{Q}_{\tilde{i}})$ fields, we assign any arbitrary R-charges satisfying $R[S_{i\tilde{i}}]+R[Q_{i}]+R[\tilde{Q}_{\tilde{i}}]=2$ modulo 6 with $R[Q_{i}]+R[\tilde{Q}_{\tilde{i}}]$ an even number. We anticipate the formation of the meson field $M_{i\tilde{i}}\sim Q_{i}\tilde{Q}_{\tilde{i}}$ in the low energy confined phase of $SU(4)$ and the meson field should not break $Z_{2R}$.} From Eq.~(\ref{eq:Rcharge}), we obtain $R[P]+R[\tilde{P}]-2=(3k-4)\times(2/5)$ which implies that $R[P]+R[\tilde{P}]$ can never be 2 modulo 6.\footnote{To see this, we may assume that $(3k-4)\times(2/5)$ is an integer multiple of 6. Then we encounter an inconsistent conclusion that $k$ cannot be an integer.} In regard to the mass term of $(P_{a},\tilde{P}_{a})$ in Eq.~(\ref{eq:superpotential1}), this tells us that either $Z_{6R}$ is explicitly broken at an energy scale above $m_{P}$ or $m_{P}$ needs to be understood as a spurion field with a proper $R$-charge. In our work, we consider the later case with the choice of $R[P]+R[\tilde{P}]=4$, i.e. $k=3$. 

Specifying $m_{P}$ as a spurion field, $m_{P}$ may stem from condensation of a field coupled to $(P_{a},\tilde{P}_{a})$. For that purpose, we introduce a chiral superfield $\Phi$ with R-charge 2 whose condensation induces the spontaneous R-symmetry breaking. In the next section, we shall make an explicit explanation how introducing $\Phi$ can explain $\mathcal{O}(100){\rm TeV}$ messenger mass. Before discussing $m_{P}$, now we discuss a R-symmetry breaking scale in what follows.

\begin{table}[t]
\centering
\begin{tabular}{|c||c|c|c|c|c|c|c|} \hline
 & $5^{*}$ & $10$ & $H_{u}$ & $H_{d}$ & $P$ & $\tilde{P}$ & $\Phi$ \\
\hline
$R$-charge      &  3  &  -1 & -2 & 0 & 2 & 2 & 2 \\
\hline
\end{tabular}
\caption{R-charge assignment. For denoting quarks and leptons in the SM, we borrow their representations in a GUT model with the gauge group $SU(5)_{\rm GUT}$. $H_{u}$ ($H_{u}$) is the chiral superfield for the up (down) type Higgs. For $S_{i\tilde{i}}$ and $(Q_{i},\tilde{Q}_{\tilde{i}})$ fields, we assign any arbitrary R-charges satisfying $R[S_{i\tilde{i}}]+R[Q_{i}]+R[\tilde{Q}_{\tilde{i}}]=2$ modulo 6 with $R[Q_{i}]+R[\tilde{Q}_{\tilde{i}}]$ an even number. For $(P_{a},\tilde{P}_{a})$, another R-charge assignment is possible as far as $R[P]+R[\tilde{P}]=4$ is satisfied.}
\label{table:qn} 
\end{table}

We may assume that $\Phi$ is a low energy degree of freedom of a hidden strong dynamics and thus has the following self-interaction terms in the superpotential
\beq
W\supset-\xi\Phi+\kappa_{1}(4\pi)^{2}\frac{\Phi^{4}}{M_{P}}+...\,,
\label{eq:superpotential2}
\eeq
where $\xi$ is a dimensionful parameter and the ellipsis stands for the higher order terms. The factor $(4\pi)^{2}$ is present in the second term of Eq.~(\ref{eq:superpotential2}) on account of the assumed strong dynamics from which $\Phi$ results~\cite{Luty:1997fk,Cohen:1997rt}. After condensation of $\Phi$, terms in Eq.~(\ref{eq:superpotential2}) are expected to produce constant terms in the superpotential. For determining a constant term in $W$, there are two possibilities depending on which is greater among the two terms in Eq.~(\ref{eq:superpotential2}).

If the first term is greater than the second one, we expect that $Z_{6R}$-breaking scale is equal to $m_{3/2}\sim\mathcal{O}(1){\rm eV}$ since the natural scale for $\sqrt{|\xi|}$ is the Planck scale without any additional symmetry to suppress the first term. This possibility, however, predicts an inconsistent very low SUSY-breaking scale via $m_{P}\sim\sqrt{4\pi F_{S}}$ and Eq.~(\ref{eq:Pmass}). Therefore, we consider the case where the second term is greater than the first term in Eq.~(\ref{eq:superpotential2}), assuming a symmetry suppressing the first term, e.g. a discrete $Z_{8}$ under which $\Phi$ and the spurion $\xi$ have the charge 2 and -2 respectively.\footnote{We checked the presence of a consistent charge assignment under $Z_{8}$.} By comparing the dominating second term in Eq.~(\ref{eq:superpotential2}) to $W_{0}=m_{3/2}M_{P}^{2}$, we find that $m_{3/2}=\mathcal{O}(1){\rm eV}$ requires $\langle\phi\rangle\simeq10^{11}{\rm GeV}$ with $\kappa_{1}=\mathcal{O}(1)$ where $\phi$ is the scalar component of $\Phi$.  

We end this section by commenting on the contribution to the mixed anomaly of $Z_{6R}\otimes SU(2)_{L}^{2}$ and $Z_{6R}\otimes SU(3)_{c}^{2}$ made by $(P_{a},\tilde{P}_{a})$ fields. $(P_{a},\tilde{P}_{a})$ are the only non-MSSM fields charged under $SU(2)_{L}$ and $SU(3)_{c}$ and thus their presence spoils the cancellation of the mixed anomaly of $Z_{6R}\otimes SU(2)_{L}^{2}$ and $Z_{6R}\otimes SU(3)_{c}^{2}$. For each anomaly, the contribution from $(P_{a},\tilde{P}_{a})$ is identically $2(R[P]+R[\tilde{P}]-2)=4$. For completeness of the model, it suffices to introduce a pair of new chiral superfields ($P',\tilde{P}'$) that transform as ${\bf 5}$ and ${\bf 5}^{*}$ under $SU(5)_{\rm GUT}$ and serve as singlets under $SU(4)$. Their R-charges satisfy the condition, $R(P')+R(\tilde{P}')=0$, 
to guarantee vanishing $Z_{6R}\otimes SU(2)_{L}^{2}$ and $Z_{6R}\otimes SU(3)_{c}^{2}$ anomalies. Then, $(P',\tilde{P}')$ fields obtain the mass as heavy as $10^{11}{\rm GeV}$ from the operator $\sim\Phi P'\tilde{P}'$ whose presence does not cause any harm.

\section{Charms of the model}
\label{sec:result}
\subsection{Messenger mass}
Since $\Phi$ is assigned R-charge 2, the mass term in Eq.~(\ref{eq:superpotential1}) can be originated from 
\beq
W\supset\kappa_{2}\frac{\Phi^{2}}{M_{P}}P_{a}\tilde{P}_{a}\,,
\label{eq:Pmass}
\eeq
where $\kappa_{2}$ is a dimensionless coupling. Thus with $\kappa_{2}=\mathcal{O}(1)$, $(P_{a},\tilde{P}_{a})$ fields acquire the mass $m_{P}=\mathcal{O}(1){\rm TeV}$ at the energy scale around $\langle\phi\rangle\simeq10^{11}{\rm GeV}$ when $Z_{6R}$ gets spontaneously broken. On the other hand, at an energy scale $E$ lower than $\langle\phi\rangle\simeq10^{11}{\rm GeV}$, we have~\cite{Ibe:2007wp} 
\beq
m_{P}(E)=\frac{m_{P}(M_{\rm IRFP})}{Z_{P}(E)}=\left(\frac{E}{M_{\rm IRFP}}\right)^{\gamma_{P*}}m_{P}(M_{\rm IRFP})\,,
\label{eq:mPE}
\eeq
where $Z_{P}(E)$ and $\gamma_{P*}$ are the wavefunction renormalization constant and the anomalous dimension at the IRFP for $(P_{a},\tilde{P}_{a})$ respectively. In addition, $M_{\rm IRFP}$ is the energy scale at which the UV to IR evolution of $g_{4}$ reaches its IRFP. With $\gamma_{P*}\simeq-0.6$~\cite{Izawa:2009nz,Yanagida:2010wf}, we see that $m_{P}=\mathcal{O}(100){\rm TeV}$ can be obtained at the messenger mass scale provided we choose $M_{\rm IRFP}\simeq10^{9}{\rm GeV}$. 

In summary, $\Phi$ condensation imposes $\mathcal{O}(1){\rm TeV}$ mass to the messengers on the spontaneous $Z_{6R}$-breaking. However, thanks to the non-trivial large anomalous dimension of $(P_{a},\tilde{P}_{a})$ fields which is attributable to the strong dynamics of $SU(4)$, $m_{P}$ at the messenger threshold becomes as large as $\mathcal{O}(100){\rm TeV}$.   

\subsection{$\mu$-term and $B$-term}
\label{sec:muB}
Referring to the R-charge assignment of the model in Table.~\ref{table:qn}, we see that the $\mu$-term can arise from 
\beq
W\supset\kappa_{3}\frac{\Phi^{2}}{M_{P}}H_{u}H_{d}\,,
\label{eq:muterm}
\eeq
where $\kappa_{3}$ is a dimensionless parameter. Given $\langle\phi\rangle\simeq10^{11}{\rm GeV}$, the model produces the Higgsino mass $\mu=\mathcal{O}(1){\rm TeV}$ for $\kappa_{3}=\mathcal{O}(1)$.

On the other hand, the SUSY-breaking being attributed to the non-vanishing F-term of $S_{i\tilde{i}}$, we notice that the F-term of $\Phi$ should vanish since $\Phi$ is decoupled from $S_{i\tilde{i}}$. Thus the term in Eq.~(\ref{eq:muterm}) cannot generate a non-vanishing contribution to the mixing term for $H_{u}$ and $H_{d}$ Higgs bosons ($B$-term). In addition, the wavefunction renormalization of $H_{u}$ and $H_{d}$ cannot generate a significant non-zero $B$-term at the scale around $m_{P}$ since there are no $S_{i\tilde{i}}$-dependent terms of the wavefunction renormalization constants of $H_{u}$ and $H_{d}$ at one-loop level. The same argument applies for the MSSM trilinear scalar coupling terms ($A$-term) to make them suppressed. Note that the suppressed $A$ and $B$ terms are favorable for resolving the CP problem appearing in the MSSM.

\begin{figure}[htp]
\centering
%\hspace*{-5mm}
\includegraphics[scale=0.6]{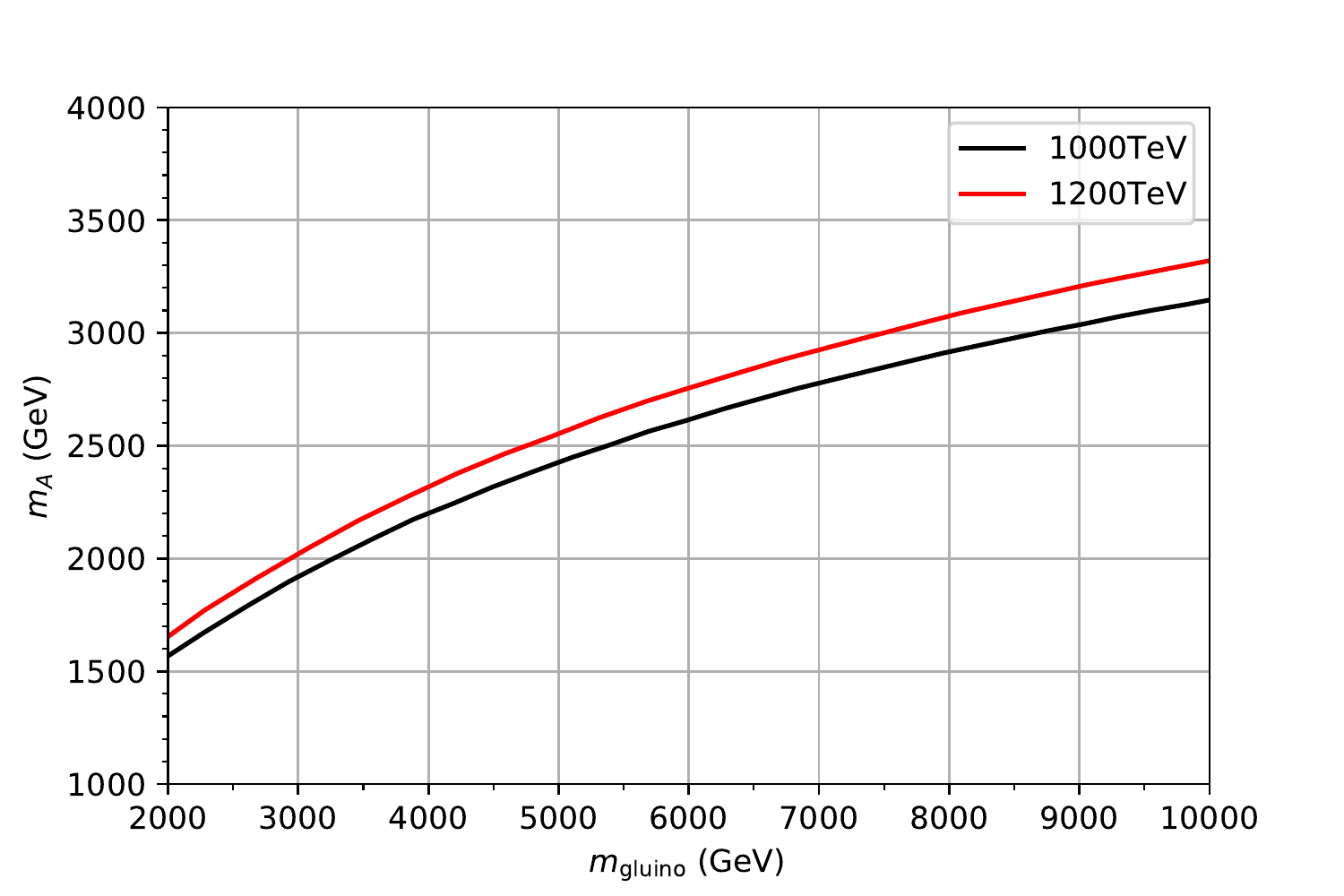}
\includegraphics[scale=0.6]{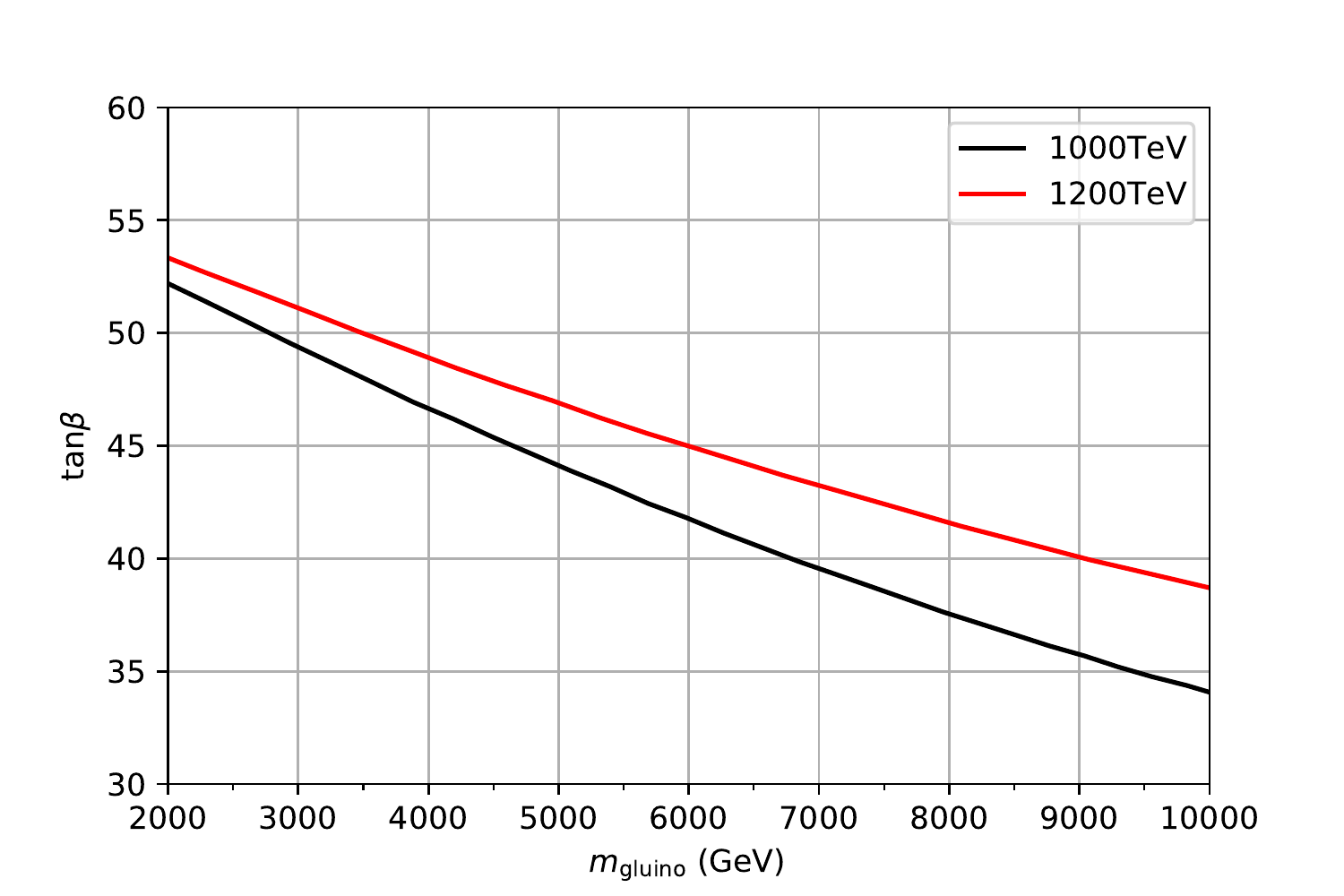}
\caption{The plots of the CP odd Higgs mass ($m_{A}$) and $\tan\beta$ as functions of the physical gluino mass. We take $\mu<0$. The black lines and red lines correspond to $\Lambda_{\rm eff}^S=1000$ and 1200\,TeV, respectively.}
%\vspace*{-1.5mm}
\label{fig:1}
\end{figure}

\begin{figure}[htp]
\centering
%\hspace*{-5mm}
\includegraphics[scale=0.6]{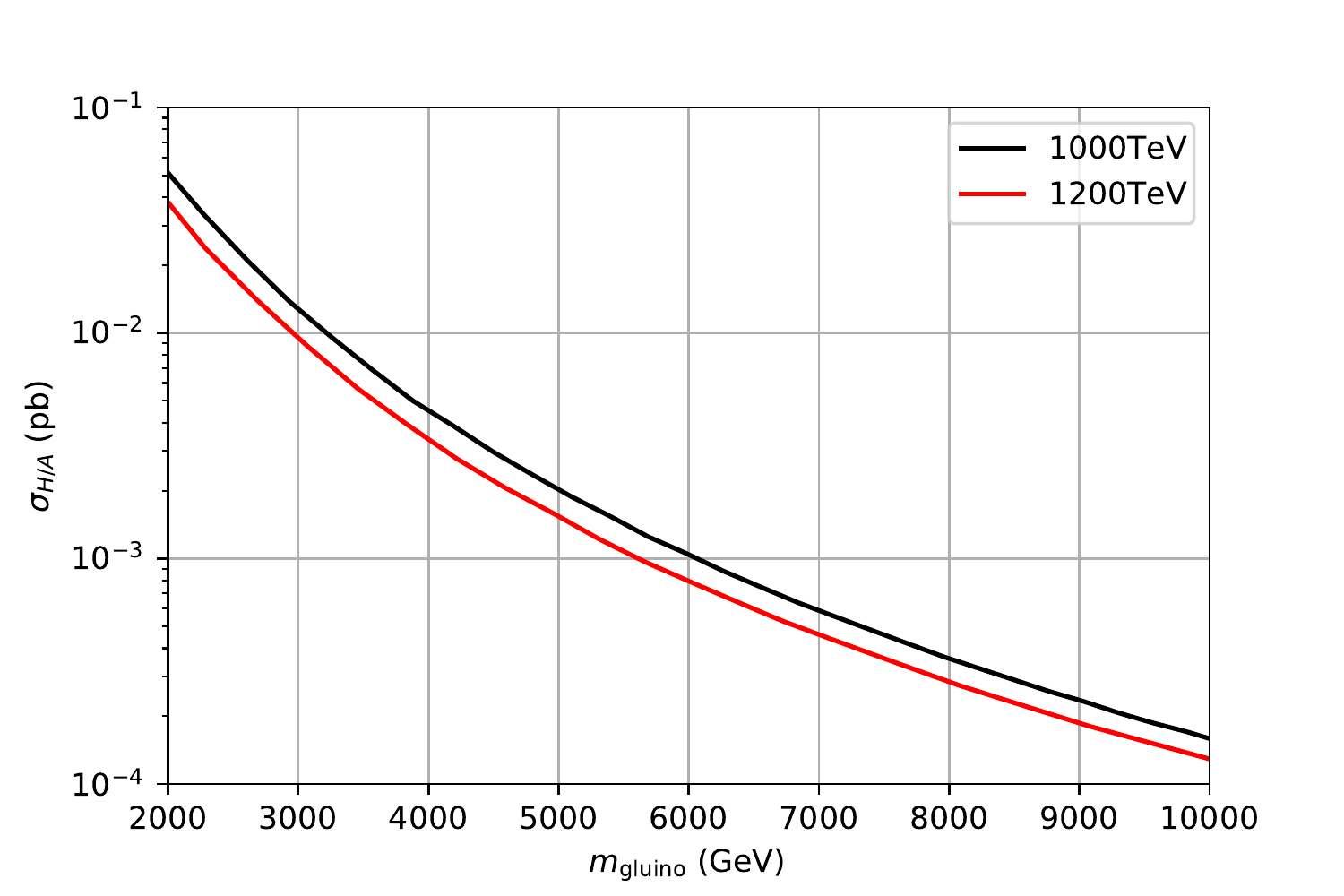}
\caption{The b-associated production cross-section of $H/A$ as a function of the physical gluino mass. The black lines and red lines correspond to $\Lambda_{\rm eff}^S=1000$ and 1200\,TeV, respectively.}
%\vspace*{-1.5mm}
\label{fig:2}
\end{figure}

Given the suppressed $B$-term at the messenger mass scale, the model is expected to result in a small value of $B$-parameter and a large ${\rm tan}\beta$ at the electroweak scale~\cite{Rattazzi:1996fb}. With this large $\tan\beta$, the masses of the second CP-even Higgs ($H$) and the CP-odd Higgs ($A$) can be as small as 2-3\,TeV as shown in Ref.~\cite{Choi:2020wdq}. Figure~\ref{fig:1} shows the predicted $m_A (\approx m_H)$ and $\tan\beta$ as functions of the physical gluino mass for $\Lambda_{\rm eff}^S= 1000$\,TeV and 1200\,TeV,
where $(\Lambda_{\rm eff}^S)^2 \sim N_{\rm eff}\Lambda_4^{4}/m_{P}^{2}$.
In the plots, we estimate the scalar masses as
\beq
m_{\rm sfermion}^{2} \simeq 2 C_{2,{\rm SM}}({\ytableausetup{textmode, centertableaux, boxsize=0.6em}
\begin{ytableau}
 \\
\end{ytableau}}) \left(\frac{\alpha_{\rm SM}}{4\pi}\right)^2 (\Lambda_{\rm eff}^S)^2.
\eeq
We take the messenger scale to be 200 TeV, where the non-zero scalar masses and gaugino masses are given as boundary conditions of renormalization group equations of MSSM. In the plots, we assume a GUT relation among the gaugino masses, i.e. $m_{\rm bino}:m_{\rm wino}:m_{\rm gluino}=g_1^2:g_2^2:g_3^2$ as in the minimal gauge mediation, where $g_1$, $g_2$ and $g_3$ are gauge coupling constants of $U(1)_Y$, $SU(2)_L$ and $SU(3)_c$, respectively. 
The mass spectrum of SUSY particles is calculated using {\tt SOFTSUSY 4.1.9}~\cite{Allanach:2001kg}. The lightest CP-even Higgs mass computed by {\tt FeynHiggs 2.18.0}~\cite{Heinemeyer:1998yj,Heinemeyer:1998np,Degrassi:2002fi,Frank:2006yh,Hahn:2013ria,Bahl:2016brp,Bahl:2017aev,Bahl:2018qog} is 124.5 GeV (125.3 GeV) for $\Lambda_{\rm eff}^S =1000$\,TeV (1200\,TeV) and $m_{\rm gluino}=3$\,TeV.
In Fig.~\ref{fig:2}, we show the b-associated production cross-section of $H/A$.
The cross section is computed using {\tt SusHi} package \cite{Harlander:2012pb,Harlander:2016hcx}. The Higgs masses, $m_A$ and $m_H$, are smaller than 3\,TeV for $m_{\rm gluino}<8$\,TeV, with the sizable production cross section. These Higgs bosons may be tested at 14-TeV LHC experiments~\cite{ATL-PHYS-PUB-2018-050}.

\subsection{Understanding the electroweak scale}
As was discussed in the previous subsection, the model is featured by a large ${\rm tan}\beta$ at the electroweak scale. Because of this, one of the electroweak symmetry breaking (EWSB) conditions yields the relation
\beq
\frac{m_{Z}^{2}}{2}\simeq-m_{H_{u}}^{2}-\mu^{2}\,,
\label{eq:mZ}
\eeq
where $m_{Z}$, $m_{H_{u}}$ and $\mu$ are the masses of Z-boson, the up-type Higgs boson and the Higgsino at the electroweak scale. In the GMSB models with a large ${\rm tan}\beta$ at the electroweak scale, there is no reason for $\mu^{2}$ to be of the same order as $-m_{H_{u}}^{2}$, which makes the fine-tuned cancellation between the two to produce $m_{Z}^{2}$ of the smaller order of magnitude incomprehensible.

To put it another way, it is challenging to understand why the uncorrelated parameters can be of the same order to produce a smaller scale , i.e. the electroweak scale. In the strongly interacting conformal gauge mediation model, however, we realize that the relation $\mathcal{O}(|-m_{H_{u}}^{2}|)=\mathcal{O}(|\mu^{2}|)$ which is necessary for Eq.~(\ref{eq:mZ}) is a natural consequence of the model and the tiny cosmological constant rather than the conspiracy between parameters. Below we demonstrate why this is the case by expressing $-m_{H_{u}}^{2}$ and $\mu^{2}$ in terms of a common parameter and showing the resulting expressions are very comparable.

Since $m_{H_u}^2$ is dominantly determined by stop loops,  
it can be related to $m_{P}$ via
\beq
-m_{H_{u}}^{2}\sim 
\frac{3}{4\pi^2}y_{t}^{2}m_{\tilde{t}}^{2} \log\frac{m_P}{m_{\tilde t}}
\sim
0.1 m_{\tilde{t}}^{2}\sim\left(\frac{\alpha_{3}}{4\pi}\right)^{2}m_{P}^{2}\,,
\label{eq:stopmass}
\eeq
where $\alpha_{3}=g_{3}^{2}/(4\pi)$, $y_{t}$ is the top Yukawa coupling and $m_{\tilde{t}}$ is the stop mass.

Meanwhile, $\mu$-parameter can be related to $m_{P}$ by the cosmological constant as shown by what follows. $\langle\phi\rangle$ determines the constant term in the superpotential $W_{0}$ throughout Eq.~(\ref{eq:superpotential2}) and thus the vanishingly small cosmological constant demands 
\beq
|F_{S}|\simeq\sqrt{3}\kappa_{1}(4\pi)^{2}\frac{\langle\phi\rangle^{4}}{M_{P}^{2}}=\sqrt{3}\frac{\kappa_{1}}{\kappa_{3}^{2}}(4\pi)^{2}\mu^{2}\,,
\label{eq:FS}
\eeq
where $F_{S}$ is the F-term of $S$-field. For the last equality, we used $\mu=\kappa_{3}\langle\phi\rangle^{2}/M_{P}$ which is indicated by the fact that the $\mu$-term is generated from the operator shown in Eq.~(\ref{eq:muterm}). Finally, combined with Eq.~(\ref{eq:FS}), the relation $F_{S}\simeq m_{P}^{2}/(4\pi)$ yields 
\beq
\mu\simeq\frac{\kappa_{3}}{(4\times\sqrt{3}\times(4\pi)^{3}\times\kappa_{1})^{1/2}}m_{P}\,.
\label{eq:mumP}
\eeq

Now taking $\alpha_{3}\simeq1/10$ in Eq.~(\ref{eq:stopmass}) and $\kappa_{1},\kappa_{3}=\mathcal{O}(1)$ in Eq.~(\ref{eq:mumP}), we encounter the notable consequence of the model: both of $-m_{H_{u}}^{2}$ and $\mu^{2}$ are parametrized by the common parameter $m_{P}$ and very close to each other as a result of the model's structure.

Therefore, the closeness of independent parameters in their magnitudes can be understood natural in the strongly interacting conformal gauge mediation model with the aid of the empirical condition for the vanishing cosmological constant. And ultimately this helps us understand the separation between the EWSB scale and the soft SUSY-breaking mass scale. 

\subsection{Cosmologically safe $\Phi$}
The chiral superfield $\Phi$ was newly introduced in Sec.~\ref{sec:Phifield} as the field whose condensation is responsible for the spontaneous breaking of $Z_{6R}$ down to $Z_{2R}$. This means that its condensation produces three distinct degenerate vacua. Hence, the model predicts the formation of domain walls at the time when $Z_{6R}$ breaks down and reduces to $Z_{2R}$. To avoid the situation where the energy budget of the universe is dominated by that of these domain walls, it is required for the model to assume that the breaking of $Z_{6R}$ takes place prior to the end of the inflation.

Yet, the simple assumption for breaking of $Z_{6R}$ during the inflation era does not fully resolve the potential domain wall problem. Even if the scalar component ($\phi$) of $\Phi$ sits at a global minimum of its potential during the inflation, it is still probable to have the symmetry restoration at the post inflation era in the case where the field fluctuation is too large. When the inflaton field oscillates around the origin of its field space at the end of the inflation, $\phi$ may begin its oscillation as well and this oscillation of $\phi$ can induce the growth of its own fluctuation via parametric resonance depending on a form of the potential. Then given Eq.~(\ref{eq:superpotential2}), one may wonder whether the model suffers from such a dangerous symmetry restoration or not.

To see whether the restoration of $Z_{6R}$ after the inflation is likely to happen, here we examine the potential of $\phi$ in the early universe. During the inflation, we may have the Hubble induced mass term for $\phi$ due to its coupling to the inflaton field. We assume a sufficiently large Hubble expansion rate during the inflation ($H_{\rm inf}$) so that the Hubble induced mass term and $\phi^{6}$-term in $V(\phi)$ dominate over the cubic term. The reason for this assumption is to become clear soon. Accordingly, we have the following effective potential of $\phi$ during the inflation
\beq
V(\phi)_{\rm eff,inf}=-c_{H}H_{\rm inf}^{2}\phi^{2}+16\kappa_{1}^{2}(4\pi)^{4}\frac{\phi^{6}}{M_{P}^{2}}\,,
\label{eq:Veff}
\eeq
where the negative Hubble induced mass is necessary to break $Z_{6R}$ during the inflation and $c_{H}>0$ is a dimensionless coupling. Assuming $c_{H},\kappa_{1}=\mathcal{O}(1)$, application of $\partial V(\phi)_{\rm eff}/\partial\phi=0$ to Eq.~(\ref{eq:Veff}) yields 
\beq
\langle|\phi_{\rm inf}|\rangle\simeq\frac{\sqrt{H_{\rm inf}M_{P}}}{48^{1/4}(4\pi)
}\simeq5\times10^{14}\times\sqrt{\frac{H_{\rm inf}}{10^{14}{\rm GeV}}}{\rm GeV}\,.
\label{eq:phiinf}
\eeq
 
After the inflation ends, while the inflaton field ($\sigma$) oscillates around a global minimum of its potential $V(\sigma)_{\rm inf}$, the effective potential of $\phi$ will be modified to 
\beq
V(\phi)_{\rm eff,osc}=-c_{H}\frac{V(\sigma)_{\rm inf}}{M_{P}^{2}}\phi^{2}+16\kappa_{1}^{2}(4\pi)^{4}\frac{\phi^{6}}{M_{P}^{2}}\,.
\label{eq:Veff2}
\eeq
It is expected that following the oscillation of $\sigma$, $\phi$ begins its oscillation which can possibly drive the growth of the field fluctuation. Now regarding this point, of particular interest in Eq.~(\ref{eq:Veff2}) is the presence of $\phi^{6}$ (sextet) term. We notice that the growth of the fluctuation of $\phi$ during $\phi$-oscillation due to parametric resonance~\cite{Kofman:1994rk,Kofman:1997yn} is expected to be inefficient for the sextet potential, which was both analytically and numerically confirmed in Ref.~\cite{Harigaya:2015hha}.\footnote{Note that for the quartic (or cubic) potential case, there exist both of negative friction on the motion of $\phi$ and the efficient parametric resonance of field. This allows for the symmetry restoration at the post inflationary era for the quartic (or cubic) potential case~\cite{Harigaya:2015hha}.}

In spite of this, one may still wonder that at a certain time during the oscillation of $\phi$, the cubic term of the potential due to Eq.~(\ref{eq:superpotential2}) can become greater than $\phi^{6}$-term so that the growth of the fluctuation of $\phi$ due to the parametric resonance becomes efficient. Even in this case, however, we see that the restoration of $Z_{6R}$ is avoided since the field value ($\phi_{\star}$) at this transition time (the cubic term becomes as large as $\phi^{6}$-term) satisfies~\cite{Kawasaki:2013iha}
\beq
\phi_{\star}=\left(\frac{\xi M_{P}}{2\kappa_{1}(4\pi)^{2}}\right)^{1/3} < 10^{4}\times\langle\phi\rangle_{0}\simeq10^{15}{\rm GeV}\,,
\label{eq:phistar}
\eeq
where $\langle\phi\rangle_{0}\simeq10^{11}{\rm GeV}$ is the required VEV of $\phi$ at the post inflationary era for the breaking of $Z_{6R}$. Note that $\xi M_{P}$ can be at most $10^{33}{\rm GeV}^{3}$ to make the first term in Eq.~(\ref{eq:superpotential2}) smaller than the second one. So Eq.~(\ref{eq:phistar}) is indeed easily satisfied. Therefore, basically Eq.~(\ref{eq:Veff2}) and Eq.~(\ref{eq:phistar}) preclude our concern for the potential restoration of $Z_{6R}$ symmetry to cause the domain wall problem at the post inflationary era.

We end this subsection by pointing out a lower bound on $H_{\rm inf}$ to make our discussion made thus far valid. Recall that Eq.~(\ref{eq:Veff}) was obtained by assuming a large enough $H_{\rm inf}$ to make the Hubble induced mass term and the sextex term much greater than the cubic term. We notice that this assumption is justified when $\langle|\phi_{\rm inf}|\rangle$ in Eq.~(\ref{eq:phiinf}) is greater than $\phi_{\star}$ in Eq.~(\ref{eq:phistar}). Thus from this requirement, we obtain the following lower bound on $H_{\rm inf}$
\beq
\langle|\phi_{\rm inf}|\rangle>\phi_{\star}\quad\rightarrow\quad H_{\rm inf}>10^{5}{\rm GeV}\,.
\eeq

\subsection{Dark Matter}
As one of the essential ingredients of the model, the gravitino is assumed to have the mass as light as $\mathcal{O}(1){\rm eV}$. As such, it is relativistic when produced from the decay of MSSM particles or scattering events among MSSM particles. Therefore, it contributes to the current dark matter (DM) population as a warm dark matter (WDM) as long as it is the lightest supersymmetry particle (LSP) in the model. Indeed, it is LSP in the strongly interacting conformal gauge mediation model and explains $\mathcal{O}(1)\%$ of the current DM abundance~\cite{Viel:2005qj}.\footnote{Since the fraction of DM abundance attributed to the light gravitino, $f_{3/2}=\rho_{3/2,0}/(\rho_{\rm 3/2,0}+\rho_{\rm CDM,0})$, is proportional to $m_{3/2}$, one obtains $1/10$ of the constraint on $f_{3/2}\lesssim0.12$ given in Ref.~\cite{Viel:2005qj} as the expected $f_{3/2}$ for $m_{3/2}=\mathcal{O}(1){\rm eV}$.}

On the other hand, as a stable and neutral low energy degree of freedom of $SU(4)$ gauge theory, we can have the following composite states as the candidate of the cold dark matter (CDM)~\cite{Yanagida:2010zz}\footnote{For another composite DM candidate in a low-scale gauge mediation model, see Ref.~\cite{Hamaguchi:2009db}.}
\beq
\Phi_{\rm DM}\supset QQQ\tilde{Q}^{\dagger},QQ\tilde{Q}^{\dagger}\tilde{Q}^{\dagger},Q\tilde{Q}^{\dagger}\tilde{Q}^{\dagger}\tilde{Q}^{\dagger}...\,,
\label{eq:composite}
\eeq
which is $SU(4)$ invariant via the omitted contraction with anti-symmetric $\epsilon$-tensor. As discussed in Sec.~\ref{sec:hiddenstrong}, $m_{3/2}=\mathcal{O}(1){\rm eV}$ implies the SUSY-breaking scale within $\mathcal{O}(100){\rm TeV}$. Since the mass of the above CDM candidate is expected to be around $\Lambda_{4}\simeq\sqrt{4\pi F_{S}}$, we see that it satisfies the upper bound on a thermal DM obtained from the partial wave unitarity~\cite{Griest:1989wd,Huo:2015nwa}. 

Hence, basically the strongly interacting conformal gauge mediation is the very model which naturally embodies the $\Lambda$CWDM model studied in Ref.~\cite{Viel:2005qj,Osato:2016ixc}. As was pointed out in those references, the scenario (especially $m_{3/2}$) will be tested by future probes of the matter power spectrum at scales $k=\mathcal{O}(0.01)-\mathcal{O}(0.1)h^{-1}{\rm Mpc}$ with a higher resolution.

%%%%%%%%%%%%%%%%%%%%%%%%%%%%%%%%%%%

% ==================================================================
\section{Conclusion}
\label{sec:conclusion}
In this paper, by extending the existing CGM model, we constructed a GMSB model featured by a hidden strong dynamics with the conformal phase at a high energy. Motivated by the phenomenological problems in SUSY models, we assumed a light gravitino mass scenario with $m_{3/2}=\mathcal{O}(1){\rm eV}$ (corresponding to $\mathcal{O}(100){\rm TeV}$ SUSY-breaking scale) which is allowed by the current astrophysical and the LHC constraint. For the strong dynamics, a hidden $SU(4)$ gauge symmetry with three pairs of matter fields $(Q_{i},\tilde{Q}_{\tilde{i}})$ is assumed. The matter sector of the $SU(4)$ gauge theory is extended with additional five pairs of messenger fields $(P_{a},\tilde{P}_{a})$ whose presence ensures the presence of an IRFP. As a consequence, both the SUSY-breaking scale and the messenger mass ($m_{P}$) become equal to one another up to an $\mathcal{O}(1)$ factor, making the invisible sector parametrized by a single parameter, i.e. $m_{P}$. 

On top of this, guided by the question for an origin of $m_{P}$, we introduced the R-symmetry ($Z_{6R}$) breaking field $\Phi$ with R-charge 2.   Remarkably, thanks to the nontrivial structure of the model (strong dynamics, conformal phase in a high energy regime), $m_{P}=\mathcal{O}(100){\rm TeV}$ was shown to stem from R-symmetry breaking scale given by the cosmological constant. Therefore, the model unifies origins of the SUSY-breaking scale, messenger mass and R-symmetry breaking scale with a great self-consistency although nothing enforces such a correlation in principle. We emphasize that the value of the model precisely lies at this point: the model is parametrized by a single dimensionful parameter like QCD.

We also presented unexpected charms of the model: the origin of the messenger mass, $\mathcal{O}(1){\rm TeV}$ $\mu$-term, the suppressed $B$-term at the electroweak scale, the possibility of having the light second Higgs bosons, the natural (comprehensible) electroweak scale, cosmologically safe introduction of $\Phi$ and dark matter candidates. We notice that these rich and appealing aspects of the model except the last one is totally ascribed to the introduction of R-symmetry breaking $\Phi$ field with R-charge 2 and thus new merits of our model as compared to the existing CGM. The presence of $\mathcal{O}(1)$ ambiguities inherent in values of quantities for which we estimate based on the naive dimensional analysis is considered the weak point of the theory, which we hope to be improved with a non-perturbative approach like the lattice computation.

% ==================================================================

\begin{acknowledgments}
T. T. Y. is supported in part by the China Grant for Talent Scientific Start-Up Project and the JSPS Grant-in-Aid for Scientific Research No. 16H02176, No. 17H02878, and No. 19H05810 and by World Premier International Research Center Initiative (WPI Initiative), MEXT, Japan. 

\end{acknowledgments}

% ================================================================

\bibliography{main}

\end{document}